\documentclass[twocolumn,a4paper,amsmath,amssymb,nofootinbib]{revtex4}%
\usepackage{graphicx}

\newenvironment{caixa}
 {\begin{tabular}{|c|} \hline\\[-10pt]}
 {\\[10pt] \hline \end{tabular}}

 \newcommand{\ud}{\mathrm{d}}

\begin{document}
\title{Tutorial de Eletromagnetismo}
\author{Christine C\'ordula Dantas} \email{ccdantas@iae.cta.br}
\affiliation{Divis\~ao de Materiais, Instituto de Aeron\'autica e Espa\c co (IAE), Centro T\'ecnico Aeroespacial (CTA), P\c ca. Mal. Eduardo Gomes, 50, CEP 12.228-904 - Vila das Ac\'acias - S\~ao Jos\'e dos Campos - SP - Brazil}
\date{\today}
\begin{abstract}
O presente tutorial visa cobrir os fundamentos do eletromagnetismo, de forma condensada e clara. Alguns exerc\'{\i}cios resolvidos e propostos foram inclu\'{\i}dos. Assume-se conhecimento de eletricidade b\'asica, derivadas parciais e integrais m\'ultiplas.
\end{abstract}

\maketitle

\tableofcontents


\section{Conceitos e Defini\c c\~oes Preliminares}

\subsection{Carga el\'etrica}

No sistema internacional de unidades (SI)\footnote{As unidades s\~ao representadas entre colchetes, ``[ ]"}, adotado no presente tutorial, utiliza-se o metro [m] como unidade de comprimento, o kilograma [kg] como unidade de massa, e o segundo [s] como unidade de tempo. Para o eletromagnetismo, \'e necess\'ario mais uma unidade b\'asica de medida -- para a carga el\'etrica. Esta unidade \'e de\-no\-minada coulomb [C].

\vspace{0.5cm}
\begin{caixa}
\begin{minipage}[c]{0.9\linewidth} 
\noindent \sl O coulomb \'e a carga que, quando colocada no v\'acuo, a 1 m de uma carga igual, repele-a com uma for\c ca de $8,9874 \times 10^9$ newtons [N].
\end{minipage}
\end{caixa}
\vspace{0.5cm}

{\sl Nota:} a carga do el\'etron \'e de $1,6 \times 10^{-19}$ [C].

\subsection{Campo el\'etrico, campo magn\'etico e for\c ca de Lorentz}

A qualquer regi\~ao onde uma carga el\'etrica $q$ experimenta uma for\c ca el\'etrica associa-se o conceito de {\sl campo el\'etrico}, que \'e usualmente uma fun\c c\~ao vetorial do espa\c co e do tempo. A for\c ca \'e devida à presen\c ca de outras cargas na regi\~ao.  A intensidade do campo el\'etrico num dado ponto \'e igual à for\c ca por unidade de carga colocada nesse ponto, isto \'e:
\begin{equation}
\vec{E} = {\vec{F} \over q}.
\end{equation}

Analogamente, um corpo magnetizado produz um {\sl campo magn\'etico} ao seu redor. Quando colocamos uma carga el\'etrica em repouso em um campo magn\'etico, nenhuma for\c ca \'e observada sobre a carga. Mas quando esta se movimenta em uma regi\~ao onde h\'a um campo magn\'etico, uma nova for\c ca \'e observada sobre a mesma, dada por:
\begin{equation}
\vec{F}=  q \left ( \vec{v} \times \mu_0 \vec{H} \right ).
\end{equation}
$\mu_0$ \'e uma constante dimensional chamada de {\sl permeabilidade do v\'acuo} (a ser discutida adiante), e $\vec{H}$ \'e a {\sl intensidade do campo magn\'etico}. Definiremos a {\sl densidade do fluxo magn\'etico no espa\c co livre} como:
\begin{equation}
\vec{B}^{*} = \mu_0 \vec{H}. \label{B*}
\end{equation}

Note que a intensidade do campo (ou densidade do fluxo) magn\'etico \'e tamb\'em uma fun\c c\~ao vetorial que depende das coordenadas de posi\c c\~ao e tamb\'em pode variar no tempo.

A {\sl for\c ca de Lorentz} \'e a soma vetorial das duas for\c cas estabelecidas em (1) e (2), isto \'e:
\begin{equation}
\vec{F}=  q( \vec{E} + \vec{v} \times \mu_0 \vec{H}). \label{lorentz}
\end{equation}
\vspace{0.5cm}
\begin{caixa}
\begin{minipage}[c]{0.9\linewidth} 
\noindent \sl A lei de Lorentz expressa o efeito do campo eletromagn\'etico numa carga em movimento.
\end{minipage}
\end{caixa}
\vspace{0.5cm}

\subsection{Unidades}

A lei de Lorentz define as unidades de $\vec{E}$ e de $\mu_0 \vec{H}$, considerando a  seguinte defini\c c\~ao para a unidade derivada {\sl volt} [V]:
\begin{equation}
[V] = {[kg][m]^2 \over [C][s]^2}.
\end{equation}
Das Eqs. 1 e 5, temos:
\begin{equation}
[E] = {[N] \over [C] } = {[kg][m]/[s]^2 \over [C] }= {[V] \over [m] }.
\end{equation}
E da Eq. 2, 3 e 5, temos:
\begin{equation}
[B^*] = [\mu_0 H] = {[N] \over [C][m]/[s] } = {[kg]\over [C][s] }= {[V][s] \over [m]^2 }\equiv [T],
\end{equation}
onde [T] \'e a unidade derivada {\sl Tesla}. Neste ponto, note que n\~ao podemos afirmar qual a unidade de H e de $\mu_0$ separadamente. Veremos posteriormente [c.f. Eq. \ref{ampere_int}] que a unidade derivada de H \'e [H] = [A]/[m], onde a unidade derivada {\sl Amp\`ere} para a corrente el\'etrica \'e dada por: [A] = [C]/[s]. Com esta informa\c c\~ao, e dada a Eq. 7, podemos deduzir que:
\begin{equation}
[\mu_0] = {[T]\over [A]/[m]} = {[W] \over [A][m]} = {[h]\over [m] },
\end{equation}
onde [h] \'e a unidade derivada {\sl Henry}, e [W], a unidade derivada {\sl Weber}. O valor num\'erico de $\mu_0$ \'e de $\mu_0 = 4 \pi \times 10^{-7}$ [h]/[m].

Unidades equivalentes no sistema gaussiano (ou CGS) s\~ao: para a densidade do fluxo magn\'etico, [G] = {\sl Gauss}; e para a intensidade do campo magn\'etico, [Oe] = {\sl Oersted}. Na Tab. I apresentamos a convers\~ao entre estas unidades.

\begin{table}
\caption{\label{tab-geom} Convers\~ao entre unidades SI e Gaussiano (CGS) para o campo magn\'etico.}
\begin{ruledtabular}
\begin{tabular}{|l|c|cc|}
 tipo          & campo     & CGS  & SI \\ \hline
fluxo          & $\vec{B}$ & [G]  & $10^{-4}$ [T] \\
intensidade    & $\vec{H}$ & [Oe] & ${10^{3}\over 4 \pi}$ [A]/[m]\\
\end{tabular}
\end{ruledtabular}
\end{table}

{\sl Observa\c c\~ao:} Al\'em da convers\~ao entre unidades, \'e poss\'{\i}vel converter fluxo em intensidade e vice-versa.

{\noindent \rule{57ex}{0.2ex}\\}
{\bf Exerc\'{\i}cio 1:} Seja o valor da intensidade do campo magn\'etico num ponto dado por $\vec{H} = 1900$ [Oe]. Qual o valor da densidade de fluxo magn\'etico $\vec{B}$ correspondente, em unidade [T]?

{\sl Solu\c c\~ao:} 
Primeiro, devemos converter [Oe] $\rightarrow$ [A]/[m].
$$H = 1900 [Oe] = 1900 \times {10^{3}\over 4 \pi} [A]/[m].$$
E ent\~ao converter para fluxo:
$$B [T] = \mu_0 {[T]\over [A]/[m]} \times H [A]/[m]$$
$$\Rightarrow B[T] = 4 \pi \times 10^{-7} {[T]\over [A]/[m]} \times 
1900 \times {10^{3}\over 4 \pi} [A]/[m]$$
$$\Rightarrow B =  0,19 [T]$$

Note, portanto, que para passar de intensidade em [Oe] para fluxo em [T] basta multiplicar por $10^{-4}$.

{\noindent \rule{60ex}{0.2ex}\\[1ex]}

\subsection{Densidade e corrente de carga; distribui\c c\~oes singulares de carga e de corrente}

{\sl Distribui\c c\~oes cont\'{\i}nuas de carga} podem ser descritas utilizando-se o conceito de {\sl densidade de carga}, $\rho$, que \'e uma fun\c c\~ao escalar que pode depender da posi\c c\~ao e do tempo. Seja um volume pequeno $\Delta V$ comparado \`as dimens\~oes do sistema, ent\~ao:
\begin{equation}
\rho (\vec{r},t) \equiv {\textrm{carga em} ~\Delta V\over \Delta V}
\end{equation}

{\sl Distribui\c c\~oes singulares de carga} (ie, conceitos limites de superf\'{\i}cie, linha e ponto) s\~ao definidas em termos de integrais: A {\sl carga pontual} $q$ \'e definida como:
\begin{equation}
q = \lim_{\rho \rightarrow \infty , V \rightarrow 0} \int_{V} \rho \ud {\textrm V}.
\end{equation}
A {\sl densidade linear de carga} $\lambda_l$ \'e definida como:
\begin{equation}
\lambda_l = \lim_{\rho \rightarrow \infty , A \rightarrow 0} \int_{A} \rho \ud {\textrm A}.
\end{equation}
E a {\sl densidade superficial de carga} $\sigma_s$ \'e definida como:
\begin{equation}
\sigma_s = \lim_{\rho \rightarrow \infty , h \rightarrow 0} 
\int_{-{h \over 2}}^{+{h \over 2}} \rho \ud \lambda,
\end{equation}
onde $h$ \'e a espessura de um elemento de volume de carga. Nestas equa\c c\~oes, os escalares $\ud {\textrm V}$, $\ud {\textrm A}$ e $\ud \lambda$ s\~ao, respectivamente, elemento de volume, elemento de \'area e ele\-men\-to de linha (com eixo de coordenadas paralelo ao vetor unit\'ario $\hat{n}$, que tem dire\c c\~ao perpendicular \`a superf\'{\i}cie carregada).

A {\sl densidade de corrente} $\vec{J}$ \'e definida por:
\begin{equation}
\vec{J} = \rho \vec{v}, \label{J}
\end{equation}
onde $\vec{v}$ \'e a velocidade da carga. Portanto $\vec{J}$ mede a taxa de transporte de um elemento de carga por unidade de \'area: $\vec{J}\equiv \ud q/(\ud \vec{A} \ud t)$, onde o vetor $\ud \vec{A}$ tem m\'odulo igual ao elemento de \'area $\ud {\textrm A}$ e dire\c c\~ao normal \`a superf\'{\i}cie: $\ud \vec{A} = \ud ({\textrm A} \hat{n})= \hat{n} \ud {\textrm A}$.

As {\sl distribui\c c\~oes singulares de corrente} s\~ao a {\sl corrente linear} $I$ e a {\sl densidade superficial de corrente} $\vec{K}$, definidas a seguir: 

\begin{equation}
I = \lim_{|\vec{J}| \rightarrow \infty , A \rightarrow 0}
\int_{A} \vec{J} \cdot \ud \vec A;
\end{equation}
\begin{equation}
\vec{K} = \lim_{|\vec{J}| \rightarrow \infty , h \rightarrow 0}
\int_{-{h \over 2}}^{+{h \over 2}} \vec{J} \ud \lambda.
\end{equation}

\subsection{Resistividade, Condutividade, Forma Pontual da Lei de Ohm, Lei de Joule}

Na teoria de circuitos el\'etricos, a {\sl resit\^encia} \'e dada por ({\sl Lei de Ohm}):
\begin{equation}
R = {V \over I},
\end{equation}
onde $V$ \'e a {\sl voltagem} (ver coment\'ario ap\'os a Eq. \ref{eletromotriz}) e $I$, a corrente linear. A unidade de $R$ \'e [Ohm], definida como a resist\^encia entre dois pontos de um condutor atrav\'es do qual uma corrente de $1$ Amp\`ere flui como resultado de uma diferen\c ca de potencial de $1$ Volt aplicado entre os dois pontos. 

A {\sl resistividade} \'e definida como:
\begin{equation}
\rho_R \equiv {R A \over \lambda},
\end{equation}
onde $A$ \'e a \'area e $\lambda$ o comprimento do material, tendo portanto unidades de [Ohm][m]. 

A {\sl condutividade} \'e definida como:
\begin{equation}
\sigma \equiv {1 \over \rho_R},
\end{equation}
tendo portanto unidades de [Ohm$^{-1}$][m$^{-1}$] ou [Siemens][m$^{-1}$].

A {\sl forma pontual da Lei de Ohm} \'e dada por:
\begin{equation}
\vec{J} = \sigma \vec{E}. \label{ohm}
\end{equation}

A for\c ca diferencial exercida por um campo el\'etrico para mover uma carga diferencial $\rho \ud V$ \'e dada por $\ud \vec{F} = \vec{E}\rho \ud V$, sendo o trabalho incremental $\ud W = \ud \vec{F} \cdot \ud \vec{\lambda}=\rho \ud V \vec{E} \cdot \ud \vec{\lambda}$. O incremento da pot\^encia dissipada \'e portanto:
\begin{equation}
\ud P = {\ud W \over \ud t} = \rho \ud V \vec{E} \cdot {\ud \vec{\lambda}\over\ud t}=
\rho \ud V \vec{E} \cdot {\vec{v}}.
\end{equation}
Dada a Eq. \ref{J}, notamos que:
\begin{equation}
\ud P =  \vec{E} \cdot \vec{J} \ud V.
\end{equation}
Integrando, obtemos a {\sl Lei de Joule}:
\begin{equation}
P =  \int_V \vec{E} \cdot \vec{J} \ud V = \int_V \sigma E^2 \ud V. \label{joule}
\end{equation}

\section{Leis de Maxwell no espa\c co livre}

Em oposi\c c\~ao \`a lei de Lorentz, que descreve o movimento de cargas a partir do campo eletromagn\'etico, nesta se\c c\~ao estaremos interessados nos efeitos da exis\-t\^encia de cargas e de seu movimento (ie, as ``fontes") {\sl sobre} o campo eletromagn\'etico. Isto \'e, como as fontes do campo eletromagn\'etico, expressas em termos de cargas el\'etricas e densidades de corrente, d\~ao origem ao campo el\'etrico e magn\'etico. Estes efeitos s\~ao des\-critos pelas {\sl Leis de Maxwell}, que podem ser formuladas em equa\c c\~oes integrais ou diferenciais. As leis integrais podem ser usadas para determinar os campos em configura\c c\~oes sim\'etricas de carga, mas para problemas mais gerais ou real\'{\i}sticos, \'e necess\'ario o uso das leis diferencias, aplic\'aveis a cada ponto do espa\c co.

\subsection{Leis Integrais de Maxwell no espa\c co livre}

\vspace{0.5cm}
\begin{caixa}
\begin{minipage}[c]{0.9\linewidth} 
\noindent \sl A lei de Gauss [Eq. \ref{gauss_int}] descreve como a intensidade do campo el\'etrico se relaciona com sua fonte, a densidade de carga.
\end{minipage}
\end{caixa}
\vspace{0.5cm}
\begin{equation}
\oint_{A} \epsilon_0 \vec{E} \cdot \ud \vec{A} = \int_{V} \rho \ud {\mathrm V}
\label{gauss_int}\end{equation}
A {\sl permissividade do espa\c co livre} $\epsilon_0$ \'e uma constante emp\'{\i}rica. Verifiquemos sua unidade. Primeiro, note que a unidade resultante do lado direito da Eq. \ref{gauss_int} \'e [C], a carga el\'etrica dentro do volume V, cuja \'area (fechada) superficial \'e A. No lado esquerdo, temos na integral um produto escalar entre dois vetores: a {\sl densidade do fluxo de deslocamento el\'etrico com no espa\c co livre}, definido por:
\begin{equation}
\vec{D}^{*}\equiv \epsilon_0 \vec{E}, \label{D*}
\end{equation} 
e o elemento de \'area, $\ud \vec{A}$. A unidade deste \'e [m]$^2$, e a do fluxo el\'etrico dever\'a ser, por quest\~oes dimensionais, [$D^*$] = [C]/[m]$^2$. Por\'em, sabemos que a dimens\~ao de $\vec{E}$ \'e [V]/[m], logo a de $\epsilon_0$ s\'o pode ser [C]/([V][m]), ou, usando a unidade derivada {\sl Farad} [F], [$\epsilon_0$] = [F]/[m]. O valor num\'erico de $\epsilon_0$ \'e $8.854 \times 10 ^{-12}$ [F]/[m], ou, para facilitar os c\'alculos, 
$10^{-9}/36 \pi$ [F]/[m] .

{\noindent \rule{57ex}{0.2ex}\\}
{\bf Exerc\'{\i}cio 2:} Derive a {\sl Lei de Coulomb} (lei do inverso do quadrado da dist\^ancia para duas cargas el\'etricas em repouso, $q_1$ e $q_2$, situadas a uma dist\^ancia $\vec{d}_{12} = r \hat{r}_{12}$ entre si, onde $\hat{r}_{12}$ \'e um vetor unit\'ario),
\begin{equation}
\vec{F}_{12} = {q_1 q_2\over 4 \pi \epsilon_0 r^2} \hat{r}_{12},
\end{equation}
a partir da lei integral de Gauss [Eq. \ref{gauss_int}], e da lei da for\c ca de Lorentz [Eq. \ref{lorentz}].

{\sl Solu\c c\~ao:} Seja uma carga $q_1$ na origem do sistema de coordenadas. Dada a simetria esf\'erica da distribui\c c\~ao de carga, o campo el\'etrico dever\'a depender da coordenada radial $r$ e independer das coordenadas angulares $\theta$ e $\phi$. Avaliemos ent\~ao, a integral de superf\'{\i}cie para um raio arbitr\'ario:

\begin{equation}
\oint_{A}\epsilon_0 \cdot \ud \vec{A} = 
\int_0^{\pi} \left [
\int_0^{2\pi} \epsilon_0 E_r (r \sin \theta \ud \phi) 
\right ]
 (r \ud \theta) = \epsilon_0 E_r 4\pi r^2.
\end{equation}
Como toda a carga est\'a concentrada na origem, a integral volum\'etrica [Eq. \ref{gauss_int}] fornece simplesmente a carga $q_1$. Assim,
\begin{equation}
\epsilon_0 E_r 4\pi r^2 = q_1 \Rightarrow \vec{E} = {q_1 \over  4\pi\epsilon_0 r^2}
\hat{r}. \label{Epontual}
\end{equation}
De acordo com a Eq. \ref{lorentz}, no local onde est\'a $q_2$, o campo $\vec{E}$ gerado pela carga $q_1$ imp\~oe uma for\c ca em $q_2$ dada por:
\begin{equation}
\vec{F} = q_2 \vec{E} = q_2 \left ( {q_1 \over  4\pi\epsilon_0 r^2} \hat{r} \right ).
\end{equation}
Ficando assim demonstrada a lei de Couloumb.

{\noindent \rule{57ex}{0.2ex}\\}

\vspace{0.5cm}
\begin{caixa}
\begin{minipage}[c]{0.9\linewidth} 
\noindent \sl A lei integral de Amp\`ere [Eq. \ref{ampere_int}] descreve como a intensidade do campo magn\'etico se relaciona com sua fonte, a densidade de corrente.
\end{minipage}
\end{caixa}
\vspace{0.5cm}
\begin{equation}
\oint_{C}\vec{H}\cdot \ud \vec{l} = \int_{A} \vec{J} \cdot \ud \vec{A}
+ {\ud \over \ud t} \int_{A} \epsilon_0 \vec{E} \cdot \ud \vec{A},
\label{ampere_int}
\end{equation}
onde a superf\'{\i}cie aberta $A$ \'e cercada pelo contorno de linha C, e $\ud \vec{l}$ \'e o elemento de linha do contorno C. \'E denominada de {\sl corrente de deslocamento} o termo $\int_{A} \epsilon_0 \vec{E} \cdot \ud \vec{A}$.

{\sl Observa\c c\~ao:}  Note que na lei de Amp\`ere, o campo $\vec{H}$ aparece sem ser multiplicado pela permeabilidade do v\'acuo, $\mu_0$. Assim, fica evidente pela lei de Amp\`ere que [H] = [C]/([m][s])=[A]/[m], como j\'a mencionado anteriormente.

{\noindent \rule{57ex}{0.2ex}\\}
{\bf Exerc\'{\i}cio 3:} Derive a {\sl lei de conserva\c c\~ao da carga} (ou de continuidade),
\begin{equation}
\oint_{A} \vec{J} \cdot \ud \vec{A} = - {\ud \over \ud t} \int_{V} \rho \ud 
{\mathrm V}, \label{conserv_int}
\end{equation}
a partir da lei integral de Gauss [Eq. \ref{gauss_int}], e da lei de Amp\`ere [Eq. \ref{ampere_int}].

{\sl Solu\c c\~ao:} Apliquemos a lei de Amp\`ere para uma superf\'{\i}cie fechada. Isto significa que o contorno vai se ``fechando" at\'e tender a zero (numa analogia onde a superf\'{\i}cie pode ser vista como uma bolsa de pano onde o contorno \'e uma cordinha que a fecha). Assim, a integral de contorno vai a zero e as integrais de superf\'{\i}cie abertas viram integrais de superf\'{\i}cie fechadas:
\begin{equation}
\oint_{A} \vec{J} \cdot \ud \vec{A} + {\ud \over \ud t} \oint_{A} \epsilon_0
\vec{E} \cdot \ud \vec{A} = 0.
\end{equation}
Mas, pela [Eq. \ref{gauss_int}], a integral do segundo termo da eq. acima \'e dada pela integral volum\'etrica da carga, o que ent\~ao demonstra a lei de continuidade.

{\noindent \rule{57ex}{0.2ex}\\}

{\sl Observa\c c\~ao:} As leis de Gauss [Eq. \ref{gauss_int}] e Amp\`ere  [Eq. \ref{ampere_int}] relacionam os campos \`as fontes, e estas est\~ao relacionadas pela lei de conserva\c c\~ao da carga [Eq. \ref{conserv_int}]. As pr\'oximas duas leis lidam apenas com campos.

\vspace{0.5cm}
\begin{caixa}
\begin{minipage}[c]{0.9\linewidth} 
\noindent \sl A lei integral de Faraday [Eq. \ref{faraday_int}] estabelece que a circula\c c\~ao de $\vec{E}$ no contorno C \'e determinada pela taxa de mudan\c ca temporal do fluxo magn\'etico atrav\'es da superf\'{\i}cie limitada pelo contorno.
\end{minipage}
\end{caixa}
\vspace{0.5cm}
\begin{equation}
\oint_{C}\vec{E}\cdot \ud \vec{l} = - {\ud \over \ud t}
\int_{A} \mu_0 \vec{H} \cdot \ud \vec{A}.
\label{faraday_int}
\end{equation}

{\noindent \rule{57ex}{0.2ex}\\}
{\bf Exerc\'{\i}cio 4:} Discuta o significado de uma regi\~ao onde a intensidade do campo el\'etrico n\~ao apresenta circula\c c\~ao. Que formas para o contorno C devem ser escolhidas para que a lei de Faraday seja v\'alida neste caso? Exemplifique.

{\sl Solu\c c\~ao:} Pela lei de Faraday [Eq. \ref{faraday_int}], uma regi\~ao onde a intensidade do campo el\'etrico n\~ao apresenta circula\c c\~ao, isto \'e,
\begin{equation}
\oint_{C}\vec{E}\cdot \ud \vec{l} = 0
\label{E_nocirc}
\end{equation}
significa que 
\begin{equation}
{\ud \over \ud t} \int_{A} \mu_0 \vec{H} \cdot \ud \vec{A} = 0,
\end{equation}
ou seja, a taxa de mudan\c ca do fluxo magn\'etico com o tempo \'e desprez\'{\i}vel. Esta condi\c c\~ao prevalece em sistemas quase-eletrost\'aticos. 

O mais interessante nesta situa\c c\~ao \'e o fato de que, seja qual for o contorno usado ao longo da integra\c c\~ao do campo $\vec{E}$ sem circula\c c\~ao, a Eq. \ref{E_nocirc} sempre d\'a zero. Isto significa que a integral de caminho (do campo $\vec{E}$ sem circula\c c\~ao) entre entre dois pontos arbitr\'arios independende do caminho escolhido. Chamamos de {\sl for\c ca eletromotriz} a integral entre dois pontos $a$ e $b$ como sendo:
\begin{equation}
\epsilon_{ab} = \int_a^b \vec{E} \cdot \ud \vec{l}. \label{eletromotriz}
\end{equation}
Assim, para uma regi\~ao onde a intensidade do campo el\'etrico n\~ao apresenta circula\c c\~ao, a for\c ca eletromotriz independe do caminho escolhido. Neste caso, 
a for\c ca eletromotriz \'e chamada de {\sl voltagem} entre dois pontos.

Um exemplo de uma regi\~ao onde a intensidade do campo el\'etrico n\~ao apresenta circula\c c\~ao \'e aquela entre duas placas paralelas com densidade de carga uniforme gerando um campo el\'etrico est\'atico entre as mesmas.

{\noindent \rule{57ex}{0.2ex}\\}

\vspace{0.5cm}
\begin{caixa}
\begin{minipage}[c]{0.9\linewidth} 
\noindent \sl A lei integral de Gauss para o fluxo magn\'etico [Eq. \ref{gauss_mag_int}] estabelece que o fluxo magn\'etico resultante de qualquer regi\~ao delimitada por uma superf\'{\i}cie fechada \'e zero.
\end{minipage}
\end{caixa}
\vspace{0.5cm}
\begin{equation}
\oint_{A} \mu_0 \vec{H} \cdot \ud \vec{A} = 0.
\label{gauss_mag_int}
\end{equation}

\subsection{Condi\c c\~oes de continuidade}

O que ocorre, por exemplo, com a intensidade do campo magn\'etico entre dois lados de uma superf\'{\i}cie carregada? Para lidarmos com singularidades de superf\'{\i}cie \'e necess\'ario a imposi\c c\~ao de condi\c c\~oes de contorno. Estas podem ser deduzidas diretamente das Eqs. de Maxwell. Aqui iremos apenas listar estas condi\c c\~oes.

\vspace{0.5cm}
\begin{caixa}
\begin{minipage}[c]{0.9\linewidth} 
\noindent \sl A componente normal da densidade do fluxo de deslocamento el\'etrico   \'e descont\'{\i}nua (sofre um ``salto") na presen\c ca de cargas na superf\'{\i}cie que separa duas regi\~oes ($\alpha$, $\beta$) do campo [Eq. \ref{gauss_cont}].
\end{minipage}
\end{caixa}
\vspace{0.5cm}
\begin{equation}
\hat{n} \cdot \left ( \epsilon_0 \vec{E}_{\alpha} - \epsilon_0 \vec{E}_{\beta} \right ) = \sigma_s.
\label{gauss_cont}
\end{equation}

\vspace{0.5cm}
\begin{caixa}
\begin{minipage}[c]{0.9\linewidth} 
\noindent \sl A componente tangencial da intensidade magn\'etica  \'e descont\'{\i}nua (sofre um ``salto") na presen\c ca de uma corrente superficial que separa duas regi\~oes ($\alpha$, $\beta$) do campo [Eq. \ref{ampere_cont}].
\end{minipage}
\end{caixa}
\vspace{0.5cm}
\begin{equation}
\hat{n} \times\left ( \vec{H}_{\alpha} - \vec{H}_{\beta} \right ) = \vec{K}.
\label{ampere_cont}
\end{equation}

\vspace{0.5cm}
\begin{caixa}
\begin{minipage}[c]{0.9\linewidth} 
\noindent \sl A componente tangencial da intensidade do campo el\'etrico  \'e cont\'{\i}nua na presen\c ca de uma densidade superficial de cargas que separa duas regi\~oes ($\alpha$, $\beta$) do campo [Eq. \ref{faraday_cont}].
\end{minipage}
\end{caixa}
\vspace{0.5cm}
\begin{equation}
\hat{n} \times\left ( \vec{E}_{\alpha} - \vec{E}_{\beta} \right ) = 0.
\label{faraday_cont}
\end{equation}

\vspace{0.5cm}
\begin{caixa}
\begin{minipage}[c]{0.9\linewidth} 
\noindent \sl A componente normal do fluxo magn\'etico   \'e cont\'{\i}nua entre duas regi\~oes ($\alpha$, $\beta$) do campo [Eq. \ref{gauss_mag_cont}].
\end{minipage}
\end{caixa}
\vspace{0.5cm}
\begin{equation}
\hat{n} \cdot \left ( \mu_0 \vec{H}_{\alpha} - \mu_0 \vec{H}_{\beta} \right ) = 0.
\label{gauss_mag_cont}
\end{equation}

\vspace{0.5cm}
\begin{caixa}
\begin{minipage}[c]{0.9\linewidth} 
\noindent \sl A componente normal da densidade de corrente   \'e descont\'{\i}nua entre duas regi\~oes ($\alpha$, $\beta$) do campo somente se a densidade superficial de cargas mudar com o tempo [Eq. \ref{charge_cons_cont}].
\end{minipage}
\end{caixa}
\vspace{0.5cm}
\begin{equation}
\hat{n} \cdot \left ( \vec{J}_{\alpha} - \vec{J}_{\beta} \right ) = 
- {\partial \sigma_s \over \partial t}.
\label{charge_cons_cont}
\end{equation}

{\noindent \rule{57ex}{0.2ex}\\}
{\bf Exerc\'{\i}cio 5:} Deduza todas as equa\c c\~oes de continuidade desta se\c c\~ao.

{\noindent \rule{57ex}{0.2ex}\\}

\subsection{Leis Diferenciais de Maxwell no espa\c co livre}

Utilizaremos dois teoremas \'uteis para passarmos das leis integrais para as leis diferenciais de Maxwell.

\vspace{0.5cm}
\begin{caixa}
\begin{minipage}[c]{0.9\linewidth} 
\noindent \sl O teorema integral de Gauss estabelece a corres\-pon\-d\^encia entre a integral de uma \'area fechada arbitr\'aria de um campo e a integral volum\'etrica  (li\-mi\-ta\-da por esta \'area) do divergente do mesmo campo [Eq. \ref{gauss_teo}].
\end{minipage}
\end{caixa}
\vspace{0.5cm}
\begin{equation}
\oint_{A}\vec{F} \cdot \ud \vec{A} = \int_{V} \nabla \cdot \vec{F} \ud {\textrm V}.
\label{gauss_teo}
\end{equation}

\vspace{0.5cm}
\begin{caixa}
\begin{minipage}[c]{0.9\linewidth} 
\noindent \sl O teorema integral de Stokes estabelece a corres\-pon\-d\^encia entre a integral de um contorno fechado arbitr\'ario de um campo e a integral de superf\'{\i}cie  (li\-mi\-ta\-da por este contorno) do rotacional do mesmo campo [Eq. \ref{stokes_teo}].
\end{minipage}
\end{caixa}
\vspace{0.5cm}
\begin{equation}
\oint_{C}\vec{F} \cdot \ud \vec{l} = \int_{A} \nabla \times \vec{F} 
\cdot \ud \vec{A}.
\label{stokes_teo}
\end{equation}

Apresentaremos, sem demonstra\c c\~ao, as leis diferenciais de Maxwell, que s\~ao obtidas diretamente dos teoremas das Eqs. \ref{gauss_teo} e \ref{stokes_teo}.

A lei de Gauss:
\begin{eqnarray}
\nabla \cdot \epsilon_0 \vec{E} & =  & \rho \nonumber\\
                                & {\textrm {ou}} & \nonumber\\
\nabla \cdot \vec{D}^{*}      & =  & \rho. \label{gauss_dif}
\end{eqnarray}

A lei de Amp\`ere:
\begin{eqnarray}
\nabla \times \vec{H} & = & \vec{J} + {\partial \epsilon_0 \vec{E} \over \partial t}\nonumber\\
                     & {\textrm {ou}} & \nonumber\\
\nabla \times \vec{H} & = & \vec{J} + {\partial \vec{D}^{*} \over \partial t}.
\label{ampere_dif}
\end{eqnarray}

A lei de Faraday:
\begin{eqnarray}
\nabla \times \vec{E} & = & - {\partial \mu_0 \vec{H} \over \partial t} \nonumber\\
                      & {\textrm {ou}} & \nonumber\\
\nabla \times \vec{E} & = & - {\partial \vec{B}^{*} \over \partial t}.
\label{faraday_dif}
\end{eqnarray}

A lei de Gauss para o campo magn\'etico:
\begin{eqnarray}
\nabla \cdot \mu_0 \vec{H} & = & 0 \nonumber\\
                           & {\textrm {ou}} & \nonumber\\
\nabla \cdot \vec{B}^{*} & = & 0.
\label{gauss_mag_dif}
\end{eqnarray}

Por fim, a equa\c c\~ao de continuidade (ou conserva\c c\~ao de carga), em forma diferencial:

\begin{equation}
\nabla \cdot \vec{J} = - {\partial \rho \over \partial t}.
\end{equation}

{\noindent \rule{57ex}{0.2ex}\\}
{\bf Exerc\'{\i}cio 6:} Deduza as leis diferenciais de Maxwell e a lei diferencial de continuidade de carga a partir das correspondentes leis integrais, utilizando os teoremas de Gauss e de Stokes.

{\noindent \rule{57ex}{0.2ex}\\}

\section{Leis de Maxwell em meios materiais}

\subsection{Polariza\c c\~ao}

Definamos a densidade de carga total como constitu\'{\i}da de duas componentes,
\begin{equation}
\rho = \rho_{livre} + \rho_{par},
\end{equation}
onde o primeiro termo se refere a cargas que podem se mover livremente de um s\'{\i}tio at\^omico a outro, como num condutor; e o segundo termo se refere a {\sl cargas emparelhadas}, como num material composto por \'atomos, mol\'eculas ou grupos de mol\'eculas (dom\'{\i}nios), no qual a presen\c ca de um campo el\'etrico induz {\sl momentos de dipolo el\'etricos}, definidos por:
\begin{equation}
\vec{p} = q \vec{d},
\end{equation}
onde $\vec{d}$ \'e a dist\^ancia entre um par de cargas opostas emparelhadas.

A {\sl densidade de polariza\c c\~ao} \'e definida por:
\begin{equation}
\vec{P} \equiv N q \vec{d},
\end{equation}
onde $N$ \'e o n\'umero de part\'{\i}culas polarizadas por unidade de volume. Pode-se demonstrar que:
\begin{equation}
\rho_{par} = - \nabla \cdot \vec{P}. \label{rho_par}
\end{equation}

{\noindent \rule{57ex}{0.2ex}\\}
{\bf Exerc\'{\i}cio 7:} Deduza a Eq. \ref{rho_par}.

{\sl Solu\c c\~ao:} Seja um meio contendo cargas emparelhadas. Por defini\c c\~ao, a carga total efetiva deste meio, num vo\-lu\-me arbitr\'ario $V$ \'e dada por:
\begin{equation}
Q_{par} = \int_{V} \rho_{par} \ud V.
\end{equation}
Uma segunda maneira de calcular isto \'e considerando um elemento de \'area da superf\'{\i}cie desta regi\~ao. Na vi\-zin\-han\c ca deste elemento de \'area, todos os centros de carga positiva est\~ao para fora da superf\'{\i}cie do elemento de vo\-lu\-me $\ud V = \vec{d} \cdot \ud \vec{A}$, com os centros de carga negativa para a parte de dentro da superf\'{\i}cie. Estes contribuem com uma carga efetiva negativa para $V$. Existem $N \vec{d} \cdot \ud \vec{A}$ destes centros em $\ud V$, logo a carga efetiva em $V$ \'e dada pela integral:
\begin{equation}
Q_{par} = - \oint_{A} q N \vec{d} \cdot \ud \vec{A} = - \oint_{A} \vec{P} \cdot \ud \vec{A}.
\end{equation}
Igualando a eq. acima com a anterior, e usando o teorema de Gauss [Eq. \ref{gauss_teo}], temos que:
\begin{equation}
\int_{V} \rho_{par} \ud V = - \oint_{A} \vec{P} \cdot \ud \vec{A}
= - \int_V \nabla \cdot \vec{P} \ud V.
\end{equation}
Dada a arbitrariedade do volume de integra\c c\~ao, os integrandos da eq. acima s\~ao portanto id\^enticos, o que prova a Eq. \ref{rho_par}.

{\noindent \rule{57ex}{0.2ex}\\}

Definamos a {\sl densidade do fluxo de deslocamento el\'etrico em meios materiais} como:
\begin{equation}
\vec{D} \equiv \epsilon_0\vec{E} + \vec{P} = \vec{D}^{*}  + \vec{P}. \label{D}
\end{equation}
Tomando o divergente da equa\c c\~ao acima, e levando em conta as Eqs. \ref{gauss_dif} e \ref{rho_par}, temos:
\begin{eqnarray}
\nabla \cdot \vec{D} & = & \nabla \cdot \epsilon_0\vec{E} + \nabla \cdot \vec{P} \nonumber \\
                     & = & \rho - \rho_{par} \nonumber\\
                     & = & \rho_{livre} \label{gauss_dif_mat}
\end{eqnarray}
A condi\c c\~ao de continuidade da componente normal do campo $\vec{D}$ \'e dada por:
\begin{equation}
\hat{n} \cdot \left ( \vec{D}_{\alpha} - \vec{D}_{\beta}  \right ) = \sigma_s^{livre}
\end{equation}

Se o meio material for {\sl eletricamente linear e isotr\'opico}, ent\~ao h\'a uma rela\c c\~ao linear entre $\vec{P}$ e $\vec{E}$:
\begin{equation}
\vec{P} = \epsilon_0 \chi_e \vec{E},
\end{equation}
onde $\chi_e$ \'e a {\sl susceptibilidade diel\'etrica}, donde escrevemos:
\begin{equation}
\vec{D} = \epsilon \vec{E}, \label{Dlin}
\end{equation}
com a {\sl permissividade do material} definida como:
\begin{equation}
\epsilon \equiv \epsilon_0(1 + \chi_e).
\end{equation}

\subsection{Magnetiza\c c\~ao}

As fontes do campo magn\'etico em meios materiais s\~ao dipolos magn\'eticos (mais ou menos alinhados) de el\'etrons individuais ou correntes causadas por el\'etrons ``circulantes". Faremos a seguinte identifica\c c\~ao do {\sl momento de dipolo magn\'etico} $\vec{m}$ com o momento de dipolo el\'etrico definido na se\c c\~ao anterior:
\begin{equation}
\vec{p} \leftrightarrow \mu_0 \vec{m}.
\end{equation}
Assim, em completa analogia com a se\c c\~ao anterior, temos:

A {\sl densidade de magnetiza\c c\~ao}:
\begin{equation}
\vec{M} = N \vec{m}.
\end{equation}

A {\sl densidade de carga magn\'etica}:
\begin{equation}
\rho_m \equiv - \nabla \cdot \mu_0 \vec{M}.
\end{equation}

A {\sl lei de Gauss para o campo magn\'etico} em meios materiais:
\begin{eqnarray}
\nabla \cdot \mu_0 \vec{H} & = & \rho_m \nonumber \\
                           & {\textrm {ou}}& \nonumber \\
\nabla \cdot \vec{B}^*     & = &  - \nabla \cdot \mu_0 \vec{M} \nonumber \\
                           & {\textrm {ou}}& \nonumber \\
\nabla \cdot \vec{B}       & =  & 0,\label{gauss_mag_mat1}
\end{eqnarray}
onde 
\begin{equation}
\vec{B} \equiv \mu_0\vec{H}+ \mu_0\vec{M} = \vec{B}^* + \mu_0\vec{M} \label{B}
\end{equation}
\'e a {\sl densidade de fluxo magn\'etico em meios materiais}.

A lei de Faraday em meios materiais \'e escrita como:
\begin{equation}
\nabla \times \vec{E} = - {\partial \vec{B} \over \partial t}.
\end{equation}

E a lei de Amp\`ere em meios materiais,
\begin{equation}
\nabla \times \vec{H} = \vec{J} + {\partial \vec{D} \over \partial t}
\end{equation}

Analogamente, a condi\c c\~ao de continuidade da componente normal do campo $\vec{B}$ \'e dada por:
\begin{equation}
\hat{n} \cdot \left ( \vec{B}_{\alpha} - \vec{B}_{\beta}  \right ) = 0.
\end{equation}

No caso de meios {\sl linearmente magnetizados}:
\begin{equation}
\vec{M} = \chi_m \vec{H},
\end{equation}
onde $\chi_m$ \'e a {\sl susceptibilidade magn\'etica}. Assim, podemos definir neste caso:
\begin{equation}
\vec{B} = \mu \vec{H},\label{Blin}
\end{equation}
com 
\begin{equation}
\mu \equiv \mu_0(1 + \chi_m),
\end{equation}
onde $\mu$ \'e a permeabilidade do material.

\section{Miscel\^anea}

\subsection{Teoremas \'uteis de campos vetoriais}

\subsubsection{$~$Teorema de Helmholtz}
Dados o divergente e o rotacional de um campo vetorial $\vec{F}(\vec{r})$, 
\begin{equation}
\nabla \cdot \vec{F}(\vec{r}) = D(\vec{r}),
\end{equation}
\begin{equation}
\nabla \times \vec{F}(\vec{r}) = \vec{C}(\vec{r}),
\end{equation}
podemos determin\'a-lo univocamente?

A resposta \'e {\sl sim} ({\sl Teorema de Helmholtz}), desde que, para $r \rightarrow \infty$: (i) $D(\vec{r})$ e $\vec{C}(\vec{r})$ ambos forem a zero mais rapidamente do que $1/r^2$, e (i) $\vec{F}(\vec{r})$ for a zero. Neste caso:
\begin{eqnarray}
\vec{F}(\vec{r}) & =  & \nabla \left ( 
-{1 \over 4 \pi} \int_{V^{\prime}} {D(\vec{r}^{~\prime}) \over \vec{r}-\vec{r}^{~\prime}}\ud V^{\prime}
\right )  +  \nonumber \\
                &  +  &  \nabla \times \left ( 
{1 \over 4 \pi} \int_{V^{\prime}} {\vec{C}(\vec{r}^{~\prime}) \over \vec{r}-\vec{r}^{~\prime}}\ud {V^{\prime}}
\right ).
\end{eqnarray}

\subsubsection{$~$Campos Solenoidais e Irrotacionais}
Campos {\sl solenoidais} s\~ao aqueles sem divergente: 
\begin{equation}
\nabla \cdot \vec{F} = 0;
\end{equation}
e campos {\sl irrotacionais} s\~ao aqueles sem rotacional: 
\begin{equation}
\nabla \times \vec{F} = 0.
\end{equation}

{\sl Teorema para Campos Solenoidais:} As condi\c c\~oes a seguir s\~ao equivalentes:
\begin{enumerate}
\item{$\nabla \cdot \vec{F} = 0$ em todo o espa\c co.}
\item{$\int_{A}\vec{F}\cdot\ud \vec{A}$ independe da superf\'{\i}cie.}
\item{$\oint_{A}\vec{F}\cdot\ud \vec{A}=0$ para qualquer superf\'{\i}cie fechada.}
\item{$\vec{F} = \nabla \times \vec{W}$, onde $\vec{W}$ \'e dito {\sl potencial vetorial} do campo $\vec{F}$. O campo $\vec{W}$ n\~ao \'e \'unico, pois tomando-se $\vec{W}^{~\prime} \rightarrow \vec{W}+ \nabla\Phi(\vec{r})$, temos que $\nabla \times \vec{W} = \nabla \times \vec{W}^{~\prime}$, uma vez que $\nabla \times \nabla\Phi(\vec{r})= 0$.}
\end{enumerate}

{\sl Teorema para Campos Irrotacionais:} As condi\c c\~oes a seguir s\~ao equivalentes:
\begin{enumerate}
\item{$\nabla \times \vec{F} = 0$ em todo o espa\c co.}
\item{$\int_a^b\vec{F}\cdot\ud \vec{l}$ independe do caminho.}
\item{$\oint_{C}\vec{F}\cdot\ud \vec{l}=0$ para qualquer contorno fechado.}
\item{$\vec{F} = -\nabla \Phi(\vec{r})$, onde $\Phi(\vec{r})$ \'e dito {\sl potencial escalar} do campo $\vec{F}$. O campo $\Phi(\vec{r})$ n\~ao \'e \'unico, pois tomando-se $\Phi(\vec{r})^{\prime} \rightarrow \Phi(\vec{r})+ b$, onde $b$ \'e uma constante, temos que $\nabla \Phi(\vec{r}) = \nabla \Phi(\vec{r})^{\prime}$, uma vez que $\nabla b= 0$.}
\end{enumerate}

\subsection{Evolu\c c\~ao temporal dos campos}

Suponha que num determinado instante, $t=t_0$, s\~ao fornecidos os campos para todo o espa\c co: $\vec{E}(\vec{r},t_0)$, $\vec{H}(\vec{r},t_0)$\footnote{Note que, de acordo com a lei de Gauss para o campo magn\'etico [c.f. Eq. \ref{gauss_mag_dif}], $\vec{H}$ precisa ser solenoidal e assim permanecer\'a durante toda a evolu\c c\~ao.} e $\vec{v}(\vec{r},t_0)$, onde estamos considerando uma regi\~ao com part\'{\i}culas de massa $m$, velocidade $\vec{v}$ e carga $q$ no v\'acuo. Segue que, pela lei de Gauss [Eq. \ref{gauss_dif}], a distribui\c c\~ao de densidade de carga fica determinada para este tempo:
\begin{equation}
\rho(\vec{r},t_0) = \nabla \cdot \epsilon_0 \vec{E}(\vec{r},t_0).
\end{equation}
Tamb\'em segue que a densidade de corrente tamb\'em (c.f. Eq. \ref{J}):
\begin{equation}
\vec{J}(\vec{r},t_0) = \rho(\vec{r},t_0) \vec{v}(\vec{r},t_0).
\end{equation}
As leis de Faraday [Eq. \ref{faraday_dif}] e de Amp\`ere [Eq. \ref{ampere_dif}] podem ser re-escritas como:
\begin{equation}
{\partial \vec{H}\over \partial t}\big\lvert_{(\vec{r},t_0)} = -{1\over \mu_0} (\nabla \times \vec{E}(\vec{r},t_0));
\end{equation}
\begin{equation}
{\partial \vec{E}\over \partial t}\big\lvert_{(\vec{r},t_0)} = {1\over \epsilon_0} [\nabla \times \vec{H}(\vec{r},t_0) - \vec{J}(\vec{r},t_0)].
\end{equation}
Isto significa que ${\partial \vec{H}\over \partial t}$ e ${\partial \vec{E}\over \partial t}$ tamb\'em ficam determinados em $t=t_0$. ${\ud \vec{v}\over \ud t}\big\lvert_{(\vec{r},t_0)}$ tamb\'em fica determinado de acordo com a lei de Lorentz [Eq. \ref{lorentz}]. Para um instante seguinte, $t = t_0 + \Delta t$, os campos evoluem de acordo com:
\begin{equation}
\vec{F}(\vec{r},t) = \vec{F}(\vec{r},t_0) + \Delta t {\partial \vec{F}\over \partial t}\big\lvert_{(\vec{r},t_0)},
\end{equation}
onde $\vec{F}$ simboliza qualquer um dos $\vec{E}$, $\vec{H}$ ou $\vec{v}$.

\subsection{Quasi-eletrost\'atica e quasi-magnetost\'atica}

Se desprezarmos a corrente de deslocamento lei de Amp\`ere, ${\partial\epsilon_0\vec{E}\over\partial t} \sim 0$ [c.f. Eq. \ref{ampere_dif}] ($\Rightarrow$ {\sl quasi-magnetost\'atica}), ou a indu\c c\~ao magn\'etica da lei de Faraday, ${\partial\mu_0\vec{H}\over\partial t} \sim 0$ [c.f. Eq. \ref{faraday_dif}] ($\Rightarrow$ {\sl quasi-eletrost\'atica}), quaisquer efeitos de onda eletromagn\'etica tamb\'em s\~ao desprez\'{\i}veis, visto que a onda \'e origin\'aria de um acoplamento entre aqueles dois termos.

Em casos quasi-est\'aticos, dadas as fontes em um determinado instante de tempo, os campos no mesmo instante de tempo s\~ao determinados sem rela\c c\~ao com o estado das fontes num instante anterior. Figurativamente, um retrato da distribui\c c\~ao das fontes determina a distribui\c c\~ao dos campos no mesmo instante de tempo.

\subsection{As Equa\c c\~oes de Poisson e de Laplace}

\subsubsection{A Equa\c c\~ao de Poisson Escalar}

Se $\nabla \times \vec{E} = 0$ (regime quasi-eletrost\'atico, [c.f. Eq. \ref{faraday_dif}]), temos que $\vec{E}$ pode ser escrito como (c.f. ``teorema para campos irrotacionais"):
\begin{equation}
\vec{E} = -\nabla \Phi. \label{Egrad}
\end{equation}
Tomando o divergente da eq. anterior e dada a lei de Gauss [c.f. Eq. \ref{gauss_dif}], temos que:
\begin{equation}
\nabla^2\Phi = -{\rho\over\epsilon_0}, \label{poisson}
\end{equation}
chamada {\sl Eq. de Poisson escalar}. Em problemas onde a distribui\c c\~ao de cargas \'e dada, a avalia\c c\~ao de um campo quasi-est\'atico \'e equivalente portanto \`a avalia\c c\~ao de uma sucess\~ao de campos est\'aticos.

Devido \`a linearidade da eq. de Poisson, a mesma obedece ao {\sl princ\'{\i}pio da superposi\c c\~ao}, isto \'e, dadas, por exemplo, $\rho_a$ e $\rho_b$, temos que: $\rho_a + \rho_b \Rightarrow \Phi_a + \Phi_b$.

Um volume elementar de carga na posi\c c\~ao $\vec{r}^{~\prime}$ d\'a origem a um potencial na posi\c c\~ao $\vec{r}$, cuja solu\c c\~ao da Eq. \ref{poisson} (com condi\c c\~oes de contorno apropriadas) \'e dada por:
\begin{equation}
\Phi(\vec{r}) = \int_{V^{\prime}} {\rho(\vec{r}^{~\prime})\over
4 \pi \epsilon_0 \mid \vec{r}- \vec{r}^{~\prime}\mid} \ud V^{\prime}. \label{poisson_sol}
\end{equation}

\subsubsection{A Equa\c c\~ao de Poisson Vetorial}

Dada a lei de Gauss para o campo magn\'etico [c.f. Eq. \ref{gauss_mag_dif}], $\nabla \cdot \mu_0 \vec{H} = 0$, temos que (c.f. ``teorema para campos solenoidais"):
\begin{equation}
\mu_0 \vec{H} = \nabla \times \vec{A}.
\end{equation}
Por conveni\^encia, escolhamos o {\sl calibre de Coulomb}:
\begin{equation}
\nabla \cdot \vec{A} = 0. \label{gauge}
\end{equation}
Assim, a lei de Amp\`ere [c.f. Eq. \ref{ampere_dif}] para o regime quasi-magnetost\'atico fica:
\begin{equation}
\nabla \times (\nabla \times \vec{A}) = \mu_0 \vec{J}.
\end{equation}
Usando as Eqs. \ref{form3} e \ref{gauge}, temos a {\sl eq. de Poisson vetorial}:
\begin{equation}
\nabla^2 \vec{A} = -\mu_0 \vec{J}. \label{poisson_vet}
\end{equation}
Na verdade, s\~ao tr\^es equa\c c\~oes de Poisson escalares, uma para cada componente de $\vec{A}$.
A solu\c c\~ao da equa\c c\~ao de Poisson vetorial (com condi\c c\~oes de contorno apropriadas) \'e dada por (tamb\'em obedecendo ao princ\'{\i}pio da superposi\c c\~ao):
\begin{equation}
\vec{A}(\vec{r}) = {\mu_0 \over 4 \pi} \int_{V^{\prime}} {\vec{J}(\vec{r}^{~\prime})\over
\mid \vec{r}- \vec{r}^{~\prime}\mid} \ud V^{\prime}. \label{poisson_vet_sol}
\end{equation}

{\sl Observa\c c\~ao:} Se tormarmos o divergente da lei de Amp\`ere para campos quasi-magnetost\'aticos, notamos que $\nabla \cdot (\nabla \times \vec{H}) = 0 = \nabla \cdot \vec{J}$ [c.f. Eq. \ref{form1}], portanto as distribui\c c\~oes de corrente neste caso s\~ao solenoidais.

\subsubsection{A Equa\c c\~ao de Laplace}

\'E simplesmente dada quando $\rho=0$ (n\~ao em todo espa\c co, pois neste caso ter\'{\i}amos simplesmente $V=0$ em todo espa\c co; no caso em quest\~ao estamos interessados em $\rho=0$ numa dada regi\~ao, havendo cargas em outras regi\~oes):
\begin{equation}
\nabla^2 \Phi = 0. \label{laplace}
\end{equation}
Solu\c c\~oes da Eq. de Laplace s\~ao ditas ``fun\c c\~oes harm\^onicas". Possuem as seguintes propriedades (aqui citadas para o caso 3D, com comportamento an\'alogo para 2D, 1D):
\begin{itemize}
\item{O valor de $\Phi$ num ponto $P$ \'e dado pelo valor m\'edio de $\Phi$ numa superf\'{\i}cie esf\'erica de raio $R$ centrada em $P$.}
\item{$\Phi$ n\~ao pode ter m\'aximos ou m\'{\i}nimos; valores extremos de $\Phi$ ocorrem nos contornos da regi\~ao.}
\end{itemize} 

{\sl Primeiro teorema da unicidade:} A solu\c c\~ao da Eq. \ref{laplace} em uma dada regi\~ao \'e unicamente determinada se $\Phi$ \'e uma fun\c c\~ao com valores especificados em todos os contornos da regi\~ao.

{\noindent \rule{57ex}{0.2ex}\\}
{\bf Exerc\'{\i}cio 8:} Demonstre o teorema anterior.

{\sl Solu\c c\~ao:} Imagine uma regi\~ao vazia cercada por uma superf\'{\i}cie fechada. Imagine que existam duas solu\c c\~oes diferentes para o potencial, $\Phi_1$ e $\Phi_2$, por\'em ambas possuindo os mesmos valores na superf\'{\i}cie, $\Phi_1(S) = \Phi_2(S)$. Tome $\Phi_3 = \Phi_1 - \Phi_2$. Note que: $\nabla^2 \Phi_3 = \nabla^2 \Phi_1 - \nabla^2 \Phi_2 = 0$, pois $\nabla^2 \Phi_1 = 0 = \nabla^2 \Phi_2$ (s\~ao solu\c c\~oes da Eq. de Laplace). Consequentemente, $\Phi_3$ tamb\'em obedece \`a Eq. de Laplace, e mais, tem o valor zero na superf\'{\i}cie: $\Phi_3(S) = 0$, pois $\Phi_3(S) = \Phi_1(S)-\Phi_2(S)=0$. Mas como n\~ao \'e permitido m\'aximos e m\'{\i}nimos, exceto na superf\'{\i}cie, e na mesma o valor de $\Phi_3$ \'e zero, segue que $\Phi_3=0$ em toda parte, e portanto $\Phi_1=\Phi_2$.

{\noindent \rule{57ex}{0.2ex}\\}

{\sl Corol\'ario:} O potencial $\Phi$ numa dada regi\~ao \'e univocamente determinado se: (a) a densidade de carga na regi\~ao, e (b) o valor de $\Phi$ em todos os contornos, s\~ao especificados. (A prova segue de maneira inteiramente an\'aloga ao exerc\'{\i}cio anterior).

\vspace{0.5cm}

{\sl Observa\c c\~ao:} Quando a distribui\c c\~ao de cargas \'e fornecida em {\sl todo o espa\c co}, a integral de superposi\c c\~ao [c.f. Eq. \ref{poisson_sol}] pode ser usada para determinar o potencial que satisfa\c ca a Eq. de Poisson [Eq. \ref{poisson}]. No entanto, h\'a casos onde a regi\~ao de interesse \'e {\sl limitada} por superf\'{\i}cies onde o potencial precisa satisfazer condi\c c\~oes de contorno especificadas (equipotenciais). Os teoremas de unicidade podem ser usados para se obter a solu\c c\~ao para o potencial, pois garantem que somente um potencial para as dadas condi\c c\~oes de contorno especificadas pode existir. T\'ecnicas como o ``m\'etodo das imagens" podem ser u\-sa\-das para este fim. Essencialmente, consiste em substituir o problema por outro inteiramente diferente, onde se tenta descobrir que distribui\c c\~ao de cargas, externas \`a regi\~ao de interesse, faz com que o potencial resultante gere a mesmas condi\c c\~oes de contorno do problema original. Pelo teorema da unicidade, o potencial equivalente assim encontrado tem que ser igual ao do problema original. Um segundo m\'etodo para resolver as Eqs. de Laplace e Poisson, mais direto, \'e o da ``separa\c c\~ao de vari\'aveis". \'E aplic\'avel quando o potencial $\Phi$ (ou sua derivada normal \`a superf\'{\i}cie $\partial \Phi /\partial n$) \'e especificado nos limites de uma dada regi\~ao, e deseja-se encontrar o potencial no interior desta regi\~ao.

\subsection{Propriedade dos Condutores}
\begin{itemize}
\item{$\vec{E}=0$ no interior de um condutor. Isto \'e, o campo $\vec{E}_{ind}$ gerado pelas cargas induzidas por um campo externo $\vec{E}_0$ tende a cancel\'a-lo no interior do condutor.}
\item{$\rho=0$ no interior de um condutor [via item anterior e lei de Gauss, Eq. \ref{gauss_dif}].}
\item{Quaisquer cargas excedentes residem na superf\'{\i}cie do condutor.}
\item{$\Phi =$ constante, no condutor todo; a superf\'{\i}cie de um condutor \'e sempre um equipotencial. Sejam $a$ e $b$ pontos quaisquer do condutor (no interior ou na superf\'{\i}cie do mesmo). Temos que $\Phi(b) - \Phi(a) = - \int_a^b \vec{E} \cdot \ud \vec{l} = 0 \Rightarrow \Phi(a)= \Phi(b)$.}
\item{$\vec{E}$ \'e perperdicular \`a superf\'{\i}cie do condutor, imediatamente do lado de fora do mesmo.}
\end{itemize}

{\sl Segundo teorema da unicidade:} Numa regi\~ao contendo condutores e preenchida por uma densidade de carga especificada, o campo el\'etrico \'e univocamente determinado se a carga total em cada condutor \'e dada.

\section{Ondas Eletromagn\'eticas}

\subsection{No v\'acuo}

Campos eletromagn\'eticos podem existir em regi\~oes muito distantes de suas fontes porque podem se propagar como ondas eletromagn\'eticas, origin\'arias do acoplamento entre $\vec{H}$ e $\vec{E}$. Examinemos este acoplamento availando se as eqs. de Maxwell admitem como solu\c c\~ao particular $\vec{E}$ e $\vec{H}$ perpendiculares entre si, i.e., com componentes dadas por:
\begin{equation}
E_x = f(z), E_y = 0, E_z = 0;
\end{equation}
\begin{equation}
H_x = 0, H_y = g(z), H_z = 0.
\end{equation}
Onde $f(z)$ e $g(z)$ s\~ao fun\c c\~oes quaisquer da coordenada $z$. Note que ambos campos s\~ao solenoidais, de acordo com a lei de Gauss (pois ${\partial E_x \over \partial x} = 0$; ${\partial E_y \over \partial y} = 0$; ${\partial E_z \over \partial z} = 0$, etc).  Assim, n\~ao h\'a cargas envolvidas, nem densidades de corrente. Note tamb\'em que, pela lei de Faraday [c.f. \ref{faraday_dif}]:
\begin{eqnarray}
{\partial {E}_y \over \partial x} -
{\partial {E}_x \over \partial y} & = &
- {\partial \mu_0 {H}_z \over \partial t} \nonumber \\
\Rightarrow {\partial {E}_x \over \partial y} & = & 0 \nonumber \\
\Rightarrow 0 & = & 0,
\end{eqnarray}
\begin{eqnarray}
{\partial {E}_z \over \partial y} -
{\partial {E}_y \over \partial z} & = &
- {\partial \mu_0 {H}_x \over \partial t} \nonumber \\
\Rightarrow 0 & = & 0,
\end{eqnarray}
\begin{eqnarray}
-{\partial {E}_z \over \partial x} +
{\partial  {E}_x \over \partial z} & = &
- {\partial \mu_0 {H}_y \over \partial t} \nonumber \\
\Rightarrow {\partial {E}_x \over \partial z} & = &
- {\partial \mu_0 {H}_y \over \partial t}. \label{c1}
\end{eqnarray}
Analogamente, pela lei de Amp\`ere [c.f. \ref{ampere_dif}]:
\begin{eqnarray}
{\partial {H}_y \over \partial x} -
{\partial {H}_x \over \partial y} & = &
{\partial \epsilon_0 {E}_z \over \partial t} \nonumber \\
\Rightarrow {\partial {H}_y \over \partial x} & = & 0 \nonumber \\
\Rightarrow 0 & = & 0,
\end{eqnarray}
\begin{eqnarray}
{\partial {H}_z \over \partial y} -
{\partial {H}_y \over \partial z} & = &
{\partial \epsilon_0{E}_x \over \partial t} \nonumber \\
\Rightarrow - {\partial {H}_y \over \partial z} & = &
{\partial \epsilon_0{E}_x \over \partial t} \label{c2}
\end{eqnarray}
\begin{eqnarray}
-{\partial {H}_z \over \partial x} +
{\partial {H}_x \over \partial z} & = &
{\partial \epsilon_0 {E}_y \over \partial t} \nonumber \\
\Rightarrow 0 & = & 0.
\end{eqnarray}
Tomando $\partial \over \partial z$ (Eq. \ref{c1}), temos:
\begin{equation}
{\partial ^2 {E}_x \over \partial z^2} =
- {\partial ^2 \mu_0 {H}_y \over \partial z \partial t}.\label{c3}
\end{equation}
E tomando $\partial \over \partial t$ (Eq. \ref{c2}), temos:
\begin{equation}
-{\partial ^2 {H}_y \over \partial t \partial z} =
{\partial ^2 \epsilon_0{E}_x \over \partial t^2}.\label{c4}
\end{equation}
Multiplicando a Eq. \ref{c4} por $\mu_0$ e inserindo na Eq. \ref{c3}, e notando que ${\partial ^2 {H}_y \over \partial t \partial z} = {\partial ^2 {H}_y \over \partial z \partial t} $,  temos:
\begin{equation}
{\partial ^2 {E}_x \over \partial t^2}= {1 \over \epsilon_0\mu_0}
{\partial ^2 {E}_x \over \partial z^2}, \label{onda_E}
\end{equation}
que \'e uma equa\c c\~ao de onda movendo-se na dire\c c\~ao $z$ com a {\sl velocidade da luz}:
\begin{equation}
c = {1 \over \sqrt{\epsilon_0 \mu_0}} \simeq 3 \times 10^8 ~\textrm{[m]/[s]}. \label{luz}
\end{equation}
Num procedimento semelhante, encontramos tamb\'em:
\begin{equation}
{\partial ^2 {H}_y \over \partial t^2}= {1 \over \epsilon_0\mu_0}
{\partial ^2 {H}_y \over \partial z^2}. \label{onda_H}
\end{equation}

Qual a rela\c c\~ao entre as amplitudes $E_x$ e $H_y$? Note que podemos utilizar para as fun\c c\~oes $f(z)$ e $g(z)$ formas sinusoidais, pois qualquer onda pode ser expressa como uma combina\c c\~ao linear de ondas sinusoidais, sendo esta combina\c c\~ao tamb\'em uma solu\c c\~ao para a equa\c c\~ao de onda. Logo podemos confinar nossa aten\c c\~ao para ondas sinusoidais de frequ\^encia $\omega$ e n\'umero de onda $k$. Assim,
\begin{equation}
\vec{E}(z,t) = E_x e^{i(kz - \omega t)} \hat{x},
\end{equation}
\begin{equation}
\mu_0\vec{H}(z,t) = H_y e^{i(kz - \omega t)} \hat{y}.
\end{equation}
Mas a lei de Faraday [c.f. Eqs. \ref{faraday_dif} e \ref{c1}] aplicada aos campos acima imp\~oe que:
\begin{equation}
k E_x =  \omega \mu_0 H_y \Rightarrow \mu_0 H_y = B_y^* = {1 \over c} E_x. \label{razao_EH}
\end{equation}

\subsection{Equa\c c\~ao Geral de Onda: Eq. de Helmholtz}

Vimos na se\c c\~ao anterior como obter a equa\c c\~ao de onda eletromagn\'etica no v\'acuo, assumindo solu\c c\~oes particulares para $\vec{E}$ e $\vec{H}$. Agora iremos deduzir a express\~ao geral, v\'alida para propaga\c c\~ao de ondas em meios materiais lineares, isotr\'opicos, homog\^eneos e invariantes no tempo, assumindo tamb\'em que o meio n\~ao possui cargas livres, $\nabla \cdot \vec{D} = 0$ [c.f. Eq. \ref{gauss_dif_mat}].

Aplicando o rotacional em ambos os lados da lei de Faraday [c.f. \ref{faraday_dif}], temos:
\begin{equation}
\nabla \times \left ( \nabla \times \vec{E} \right ) = - \mu 
 {\partial \over \partial t} \left ( \nabla \times \vec{H} \right ).
\end{equation}
Usando a lei de Ohm [Eq. \ref{ohm}] e inserindo a lei de Amp\`ere
[c.f. \ref{ampere_dif}] na equa\c c\~ao anterior, temos:
\begin{equation}
\nabla \times \left ( \nabla \times \vec{E} \right ) = - \mu 
 {\partial \over \partial t} \left ( 
\sigma \vec{E} + \epsilon {\partial \vec{E} \over \partial t}
\right ).
\end{equation}
Usando a rela\c c\~ao \ref{form3}, e notando que o meio \'e livre de cargas, obtemos finalmente a {\sl Equa\c c\~ao de Onda de Helmholtz} para o campo $\vec{E}$:
\begin{equation}
\nabla^2 \vec{E} = \mu \sigma {\partial \vec{E} \over \partial t} 
+ \mu \epsilon {\partial ^2 \vec{E} \over \partial t^2}. \label{helmholtz}
\end{equation}
Uma express\~ao similar pode ser obtida para o campo $\vec{H}$. Note que no v\'acuo, $\sigma = 0$, donde recuperamos a Eq. \ref{onda_E}.

\subsection{Propaga\c c\~ao, Reflex\~ao e Transmiss\~ao em Meios Lineares}
Em meios lineares (homog\^eneos, isotr\'opicos, com $\epsilon$ e $\mu$ independentes da posi\c c\~ao e dire\c c\~ao), a velocidade de propaga\c c\~ao das ondas eletromagn\'eticas \'e dada por:
\begin{equation}
\nu = {1 \over \sqrt{\epsilon \mu}} = {c \over n}, \label{vel_propag}
\end{equation}
onde $n$ \'e o \'{\i}ndice de refra\c c\~ao. Suponha que o plano $yz$ forma uma fronteira entre dois meios (1 e 2). Uma onda plana de frequ\^encia $\omega$, viajando na dire\c c\~ao $x$ se aproxima da interface pelo lado esquerdo (meio 1):
\begin{eqnarray}
\vec{E}^I(x,t) & = & E^I_0 e^{i(k_1 x - \omega t)} \hat{y} \nonumber \\
\mu_1 \vec{H}^I(x,t) & = & {1 \over \nu_1} \mu_1 H^I_0 e^{i(k_1 x - \omega t)} \hat{z},
\end{eqnarray}
gerando uma onda refletida, que viaja de volta no meio 1, 
\begin{eqnarray}
\vec{E}^R(x,t) & = & - E^R_0 e^{i(k_1 x - \omega t)} \hat{y} \nonumber \\
\mu_1 \vec{H}^R(x,t) & = & - {1 \over \nu_1} \mu_1 H^R_0 e^{i(k_1 x - \omega t)} \hat{z},
\end{eqnarray}
e uma transmitida, que atravessa para o lado direito (meio 2):
\begin{eqnarray}
\vec{E}^T(x,t) & = & E^T_0 e^{i(k_2 x - \omega t)} \hat{y} \nonumber \\
\mu_2 \vec{H}^T(x,t) & = & {1 \over \nu_2} \mu_2 H^T_0 e^{i(k_2 x - \omega t)} \hat{z}.
\end{eqnarray}
Em $x=0$, os campos da esquerda conjuntamente devem se unir aos da direita, de acordo com as condi\c c\~oes de contorno [c.f. se\c c\~ao sobre condi\c c\~oes de continuidade e dedu\c c\~ao para meios lineares]. Como os campos n\~ao possuem componentes perpendiculares \`a superf\'{\i}cie de interface, temos que estas condi\c c\~oes s\~ao apenas para as componentes tangenciais [compare com as Eqs. \ref{faraday_cont} e \ref{ampere_cont}, como refer\^encia]:
\begin{eqnarray}
\hat{n} \times\left ( \vec{E}_{1} - \vec{E}_{2} \right ) & = & 0 
\Rightarrow \vec{E}_{1}^{\parallel} = \vec{E}_{2}^{\parallel} \nonumber \\
\Rightarrow E^I_0 + E^R_0 & = & E^T_0,
\end{eqnarray}
\begin{eqnarray}
\hat{n} \times\left ( \vec{H}_{1} - \vec{H}_{2} \right ) & = &
 \vec{K} = 0  \Rightarrow \vec{H}_{1}^{\parallel} =  \vec{H}_{2}^{\parallel} \nonumber \\
\Rightarrow {1\over\mu_1}\left ( {1 \over \nu_1} E^I_0 - {1 \over \nu_1} E^R_0   \right ) & = & {1 \over \mu_2} \left ( {1 \over \nu_2} E^T_0 \right )\nonumber \\
\Rightarrow  E^I_0 - E^R_0 & = & \beta E^T_0,
\end{eqnarray}
com $\beta = (\mu_1\nu_1)/(\mu_2\nu_2)$. Como $\mu_1 \sim \mu_0 \sim \mu_2$, podemos assumir $\beta = \nu_1/\nu_2$, o que nos d\'a as solu\c c\~oes:
\begin{eqnarray}
E^R_0 =  \left ( {\nu_2 - \nu_1 \over \nu_2 + \nu_1} \right ) E^I_0 
& ; &
E^T_0 = \left ( {2 \nu_2 \over \nu_2 + \nu_1} \right ) E^I_0.
\nonumber \\
\end{eqnarray}
Como a intensidade $I$ \'e proporcional \`a amplitude da onda ao quadrado, pela express\~ao $I = {\nu \over 2} \epsilon E_0^2$, temos que o {\sl coeficiente de reflex\~ao} \'e dado por:
\begin{equation}
R = {I_R \over I_I} = \left ( {E^R_0\over E^I_0} \right )^2 =
\left ( {n_1-n_2 \over n_1 + n_2} \right )^2,
\end{equation}
e o {\sl coeficiente de transmiss\~ao}:
\begin{equation}
T = {I_T \over I_I} = {\epsilon_2 \nu_2 \over \epsilon_1 \nu_1}\left ( {E^T_0\over E^I_0} \right )^2 = {n_2 \over n_1}
\left ( {2 n_1 \over n_1 + n_2} \right )^2.
\end{equation}
Note que $R+T=1$, tal como requerido pela conserva\c c\~ao de energia.

\subsection{Constante de propaga\c c\~ao, atenua\c c\~ao, constante de fase, imped\^ancia}

Consideremos campos harm\^onicos no tempo, e.g.,
\begin{equation}
\vec{E}(x,y,z,t) = Re \left [ 
\vec{E}_0 (x,y,z) e^{i(\omega t + \phi)} \right ] 
=  Re \left [  \vec{E}_s e^{i(\omega t)} \right ],
\end{equation}
com fasor definido por $\vec{E}_s \equiv \vec{E}_0 (x,y,z) e^{i\phi}$, onde $\phi$ \'e o deslocamento de fase. Dado que ${\partial \vec{E} \over \partial t} = i \omega \vec{E}_s$ e ${\partial ^2 \vec{E} \over \partial t^2} = (i \omega)^2 \vec{E}_s$, e que as derivadas espaciais dependem apenas de $\vec{E}_s$, a equa\c c\~ao de Helmholtz nos fornece [c.f. \ref{helmholtz}]:
\begin{equation}
\nabla^2 \vec{E}_s - \gamma^2 \vec{E}_s = 0,
\end{equation}
com a {\sl constante de propaga\c c\~ao} definida por
\begin{equation}
\gamma = \sqrt { i \omega \mu \left (  \sigma + i \omega \epsilon \right )}
= \alpha + i \beta, \label{gamma}
\end{equation}
onde $\alpha$ \'e dita {\sl atenua\c c\~ao} da onda (unidades de [Nepers][m]$^{-1}$), e $\beta$ \'e a {\sl constante de fase}\footnote{Tamb\'em conhecida como {\sl n\'umero de onda}, $k = 2 \pi / \lambda$.} (unidades de [rad][m]$^{-1}$). Uma express\~ao similar \'e obtida para campos magn\'eticos,
\begin{equation}
\nabla^2 \vec{H}_s - \gamma^2 \vec{H}_s = 0.
\end{equation}
Considerando uma onda plana polarizada na dire\c c\~ao $x$ e se propagando na dire\c c\~ao $z$, i.e, $\vec{E}_s (z) = E_{x,s}(z) \hat{x}$, pode-se mostrar que a solu\c c\~ao geral para $\vec{E}_s$ \'e dada pela superposi\c c\~ao linear:
\begin{equation}
\vec{E}_s = \left ( 
E_0^+ e^{- \gamma z} + E_0^- e^{+ \gamma z} 
\right ) \hat{x}.
\end{equation}
E, pela lei de Faraday [c.f. \ref{faraday_dif}], tamb\'em nota-se facilmente que
\begin{equation}
\nabla \times \vec{E}_s = - i \omega \mu \vec{H}_s \rightarrow
\vec{H}_s= - { \nabla \times \vec{E}_s \over i \omega \mu}.
\end{equation}
Resolvendo o rotacional acima, obtemos:
\begin{equation}
\vec{H}_s = \left ( 
{\gamma E_0^+ \over i \omega \mu} e^{- \gamma z} -
{\gamma E_0^- \over i \omega \mu} e^{+ \gamma z} 
\right ) \hat{y}.
\end{equation}

A {\sl imped\^ancia intr\'{\i}nseca} do meio \'e definida como:
\begin{equation}
\eta \equiv {E_0^+ \over H_0^+} = {i \omega \mu\over\gamma}. \label{imp_razao}
\end{equation}
De acordo com a defini\c c\~ao \ref{gamma}, obtemos:
\begin{equation}
\eta = \sqrt{ {i \omega \mu \over\sigma + i \omega \epsilon} }. \label{imp}
\end{equation}

Note que, no espa\c co livre, ou v\'acuo, n\~ao h\'a cargas e a condutividade \'e nula ($\sigma=0$). Logo, a constante de propaga\c c\~ao [Eq. \ref{gamma}] fica:
\begin{equation}
\gamma = \sqrt { i \omega \mu \left (  0 + i \omega \epsilon \right )}
= i\omega \sqrt{\mu \epsilon} = \alpha + i \beta,
\end{equation}
donde $\alpha = 0$ (o sinal n\~ao atenua ao se propagar) e $\beta = \omega \sqrt{\mu \epsilon}$. Esta condi\c c\~ao vale para qualquer meio sem perdas, n\~ao somente o v\'acuo, mas como tamb\'em para um {\sl diel\'etrico perfeito}.
Note que esta express\~ao concorda com a velocidade de propaga\c c\~ao da onda dada pela Eq. \ref{vel_propag}:
\begin{equation}
\nu = {\omega \over \beta} = {1 \over \sqrt{\epsilon \mu}}.
\end{equation}
Para o v\'acuo, vimos que a velocidade de propaga\c c\~ao \'e a velocidade da luz [c.f. Eq. \ref{luz}]. 

Para um diel\'etrico perfeito e n\~ao-magn\'etico (i.e., $\mu_r = \mu/\mu_0 = 1$), temos:
\begin{equation}
\nu = {1 \over \sqrt{\epsilon \mu_0}} = {1 \over \sqrt{\epsilon_r \epsilon_0 \mu_0}} = {c \over \sqrt{\epsilon_r}}.
\end{equation}
Neste caso, a imped\^ancia intr\'inseca \'e dada por [c.f. Eq. \ref{imp}]:
\begin{equation}
\eta = \sqrt{ {i \omega \mu \over 0 + i \omega \epsilon} } = 
\sqrt{\mu \over \epsilon},
\end{equation}
ou seja, um valor real. Reescrevendo a equa\c c\~ao acima como:
\begin{equation}
\eta =  \sqrt{\mu_r \mu_0 \over \epsilon_r \epsilon_0}
= \sqrt{\mu_r \over \epsilon_r} \eta_0,
\end{equation}
obtemos a imped\^ancia intr\'inseca do espa\c co livre:
\begin{equation}
\eta_0 =  \sqrt{\mu_0 \over \epsilon_0} =
\sqrt{4 \pi \times 10 ^{-7} \rm{[H][m]}^{-1} \over 
\left ( 10^{-9} / 36 \pi \right ) \rm{[F][m]^{-1}}} = 120 \pi \rm{[Ohm]}. \label{imp_vacuo}
\end{equation}

Por fim, uma pequena observa\c c\~ao quanto a raz\~ao das amplitudes dos campos, tal como expressa pela Eq. \ref{imp_razao}. Note as seguintes f\'ormulas \'uteis:
\begin{eqnarray}
\vec{H}_s & = & {1\over\eta} \hat{u} \times \vec{E}_s \\
& {\rm e}  & \nonumber \\
\vec{E}_s & = & - \eta \hat{u} \times \vec{H}_s \\
\end{eqnarray}
onde $\hat{u}$ \'e o vetor unit\'ario na dire\c c\~ao de propaga\c c\~ao da onda. Note tamb\'em que essas equa\c c\~oes concordam com os c\'alculos anteriores, onde partimos de solu\c c\~oes particulares para os campos no espa\c co livre, uma vez que, facilmente se encontra pela Eq. \ref{imp_vacuo} que $\eta_0 = \mu_0 c$. O que nos leva a concord\^ancia entre as Eqs. \ref{razao_EH} e \ref{imp_razao}:
\begin{equation}
\mu_0 H_y = {1 \over c} E_x \rightarrow H_y = {1 \over \eta_0} E_x
\end{equation}

\subsection{Propaga\c c\~ao em meios com perdas; tangente de perdas}

Verifiquemos as express\~oes das quantidades vistas na se\c c\~ao anterior para o caso de materiais que apresentam perdas (o sinal atenua ao se propagar no meio).

\subsubsection{Diel\'etricos}
Para algumas aproxima\c c\~oes, diel\'etricos podem ser tratados como diel\'etricos perfeitos (i.e., sem perdas), por\'em todos os diel\'etricos apresentam perdas em algum grau. A natureza das perdas tem origem em dois fen\^omenos (ou uma combina\c c\~ao destes):
\begin{itemize}
\item{{\sl Perdas por condutividade finita.} O campo $\vec{E}$ gera uma corrente de condu\c c\~ao $\vec{J} = \sigma \vec{E}$ [c.f. Lei de Ohm, Eq. \ref{ohm}]. A presen\c ca de $\vec{E}$ e $\vec{J}$ gera dissipa\c c\~ao de pot\^encia (como calor) por meio da Lei de Joule [c.f. Eq. \ref{joule}]. Esta dissipa\c c\~ao de pot\^encia atenua a onda eletromagn\'etica. }
\item{{\sl Perdas por polariza\c c\~ao.} Associadas \`a energia exigida pelo campo para movimentar dipolos ``relutantes''. Este mecanismo \'e proporcional \`a frequ\^encia.}
\end{itemize}

Consideremos a permissividade complexa como:
\begin{equation}
\epsilon_c = \epsilon^{\prime} - i \epsilon^{\prime\prime},
\end{equation}
onde:
\begin{itemize}
\item{$\epsilon^{\prime}$: parte real de $\epsilon_c$, i.e., $\epsilon^{\prime} \equiv \epsilon = \epsilon_r \epsilon_0$.}
\item{$\epsilon^{\prime\prime}$: parte imagin\'aria de $\epsilon_c$, que se refere \`as perdas por polariza\c c\~ao.}
\end{itemize}

A partir das Leis de Ohm [c.f. Eq. \ref{ohm}] e de Amp\`ere [c.f. Eq. \ref{ampere_dif}], aplicadas aos fasores do campo eletromagn\'etico harm\^onico no tempo, obtemos:
\begin{equation}
\nabla \times \vec{H}_s = \sigma \vec{E}_s +
i \omega (\epsilon^{\prime} - i \epsilon^{\prime\prime})\vec{E}_s,
\end{equation}
ou
\begin{equation}
\nabla \times \vec{H}_s = \left [ 
\left (
\sigma + \omega \epsilon^{\prime\prime} 
\right )
+ i \omega \epsilon^{\prime}
\right ] \vec{E}_s, \label{ampere_mod}
\end{equation}
onde podemos considerar uma {\sl condutividade efetiva},
\begin{equation}
\sigma_{\rm{ef}} \equiv \sigma + \omega \epsilon^{\prime\prime},
\end{equation}
que inclui ambas as perdas (condutividade e polariza\c c\~ao). As f\'ormulas para a constante de propaga\c c\~ao [c.f. Eq. \ref{gamma}] e para a imped\^ancia [c.f. Eq. \ref{imp}] continuam v\'alidas, aplicando-se $\sigma \rightarrow \sigma_{\rm{ef}}$. Note que ambas quantidades s\~ao complexas neste caso, e isso implica que a onda ir\'a atenuar devido \`a $\alpha > 0$ na constante de propaga\c c\~ao, e haver\'a uma diferen\c ca de fase entre os campos $\vec{E}$ e $\vec{H}$.

Determinemos as express\~oes para $\alpha$ e $\beta$ para um diel\'etrico em geral, sem tecer considera\c c\~oes ainda sobre perdas. Re-arranjando a express\~ao \ref{gamma}, temos:
\begin{equation}
\gamma^2 = -\omega^2\mu\epsilon + i \omega\mu\sigma = 
\left ( \alpha + i \beta \right ) ^2 = 
\left ( \alpha ^2 - \beta ^2 \right ) + i 2 \alpha \beta,
\end{equation}
donde, igualando os termos reais e imagin\'arios, e resolvendo as equa\c c\~oes resultantes, temos:
\begin{equation}
\alpha = \omega \sqrt{
{\mu \epsilon \over 2} 
\left (
\sqrt{
1 + \left (
{\sigma \over \omega \epsilon}
\right ) ^2 
}
- 1
\right )
}  \label{alpha_geral}
\end{equation}
\begin{equation}
\beta = \omega \sqrt{
{\mu \epsilon \over 2} 
\left (
\sqrt{
1 + \left (
{\sigma \over \omega \epsilon}
\right ) ^2 
}
+ 1
\right )
} \label{beta_geral}
\end{equation}
Ou seja, obtivemos a atenua\c c\~ao e a constante de fase em termos dos par\^ametros constitutivos de um material diel\'etrico em geral. Para incluir efeitos de perda (caracterizados pela condutividade finita $\sigma$ e a parte imagin\'aria $\epsilon^{\prime\prime}$), mais uma vez tomamos $\sigma \rightarrow \sigma_{\rm{ef}}$ nas f\'ormulas acima. 

Note que para ondas harm\^onicas no tempo a densidade corrente de deslocamento da Lei de Amp\`ere [c.f. Eq. \ref{ampere_dif}] \'e dada por (lembrando que $\epsilon = \epsilon^{\prime}$):
$\vec{J}_{{\rm des}} \equiv \partial \epsilon^{\prime} \vec{E} / \partial t = i \omega \epsilon^{\prime} \vec{E}_s$ (uma quantidade puramente imagin\'aria), enquanto que a densidade de corrente de condu\c c\~ao efetiva [c.f. Eq. \ref{ohm}] \'e dada por $\vec{J}_{{\rm ef}} = \sigma_{{\rm ef}} \vec{E}_s$ (uma quantidade real). De acordo com a Eq. \ref{ampere_mod}, temos etn\~ao:
\begin{equation}
\nabla \times \vec{H}_s = \vec{J}_{{\rm ef}} + \vec{J}_{{\rm des}} = 
\vec{J}_{{\rm tot}} 
\end{equation}

A {\sl tangente de perdas} ($\tan \delta$) \'e definida pela raz\~ao das componentes real e imagin\'aria de $\vec{J}_{{\rm tot}}$:
\begin{equation}
\tan \delta = { {\rm Re} \left[  \vec{J}_{{\rm tot}} \right ]
\over 
{\rm Imag} \left [ \vec{J}_{{\rm tot}} \right ]
} = 
{\sigma + \omega \epsilon^{\prime \prime}\over \omega \epsilon^{\prime}}
=
{\sigma_{{\rm ef}}\over \omega \epsilon^{\prime}}.
\end{equation}
O \^angulo $\delta$ portanto fornece, no plano complexo, o \^angulo no qual  $\vec{J}_{{\rm des}}$  est\'a adiantada com rela\c c\~ao \`a $\vec{J}_{{\rm tot}}$. Note tamb\'em que a tangente de perdas varia com a frequ\^encia. Alguns casos a se considerar:

\begin{itemize}
\item{{\sl ``Bom'' diel\'etrico (perdas baixas):} $\sigma \rightarrow 0$, logo $\tan \delta \approx \epsilon^{\prime\prime}/\epsilon^{\prime}$. Ou ainda, de maneira geral, $\tan \delta << 1$, i.e., $\sigma_{{\rm ef}} / \omega \epsilon^{\prime} << 1$. Podemos utilizar a aproxima\c c\~ao $(1+x)^n \approx 1 + nx$ para $x = \sigma / \omega \epsilon $ nas Eqs. \ref{alpha_geral} e \ref{beta_geral}, obtendo f\'ormulas mais simples: $\alpha \approx {\sigma \over 2} \sqrt{\mu \over \epsilon}$ e $\beta \approx \omega \sqrt{\mu \epsilon}$.}
\item{{\sl ``Bom'' condutor:} $\sigma >> \omega \epsilon^{\prime\prime}$ (excetuando em frequ\^encias suficientemente elevadas), resultando na aproxima\c c\~ao
$\tan \delta \approx \sigma/\omega \epsilon^{\prime}$. Trataremos mais em detalhes de condutores na pr\'oxima se\c c\~ao. }
\end{itemize}

\subsubsection{Condutores}

Vimos que para condutores $\sigma >> \omega \epsilon$, o que nos fornece as seguintes aproxima\c c\~oes para $\alpha$ e $\beta$:
\begin{equation}
\alpha = \beta \approx \sqrt{\omega \mu \sigma \over 2} = \sqrt{\pi f \mu \sigma}.
\end{equation}
A imped\^ancia intr\'{\i}nseca recebe a seguinte aproxima\c c\~ao:
\begin{equation}
\eta \approx \sqrt{i\omega \mu \over \sigma}.
\end{equation}
Notando a identidade $\sqrt{i} = (1+i)/\sqrt{2}$, e aplicando a f\'ormula de Euler [c.f. Eq. \ref{euler}], a aproxima\c c\~ao anterior pode ser re-escrita como:
\begin{equation}
\eta \approx \sqrt{\omega \mu \over 2\sigma}(1+i) = 
\sqrt{\omega \mu \over \sigma}e^{i 45^{\rm o}}
= \sqrt{2} {\alpha \over \omega}e^{i 45^{\rm o}}.
\end{equation}
O campo magn\'etico se encontra defasado em rela\c c\~ao ao campo el\'etrico em $45^{\rm o}$. Uma consequ\^encia de $\sigma$ grande \'e a redu\c c\~ao dr\'astica na velocidade de propaga\c c\~ao $\nu = \omega / \beta \approx \sqrt{(2 \omega)/(\mu \sigma)}$ e no comprimento de onda $\lambda = 2\pi / \beta \approx 2 \sqrt{\pi/(f \mu \sigma)}$. Uma grande atenua\c c\~ao significa que a maior parte da energia da onda incidente em um condutor ser\'a refletida, e os campos ter\~ao uma pequena profundidade de penetra\c c\~ao no material.

\subsection{Ondas TE, TM e TEM}

Ondas eletromagn\'eticas confinadas em um condutor cilindrico oco ({\sl guia de onda}) n\~ao s\~ao geralmente transversas, havendo componentes longitudinais. Isto ocorre de\-vi\-do \`as condi\c c\~oes de contorno no interior da parede interna do condutor. Pode-se demonstrar que, para uma onda eletromagn\'etica propagando, por exemplo, na dire\c c\~ao $x$ ao longo do condutor, as Eqs. de Maxwell conjuntamente com as condi\c c\~oes de contorno ($\vec{E}_{\parallel}=0$ e $\vec{B}^*_{\perp}=0$) geram um par de equa\c c\~oes desacopladas para as componentes longitudinais $E_x$ e $B^*_x$. Se $E_x=0$, as ondas eletromagn\'eticas s\~ao ditas {\sl TE} (``transverse electric") e se $B^*_x=0$, s\~ao ditas {\sl TM} (``transverse magnetic"). Se ambas condi\c c\~oes ocorrerem, $E_x=0$ e $B^*_x=0$, s\~ao ditas ondas {\sl TEM}. Pode-se demonstrar que ondas TEM n\~ao podem ocorrer em um guia de onda oco.

\subsection{Teorema de Poynting}

O {\sl Teorema de Poynting} afirma que a taxa de decr\'escimo da energia armazenada nos campos el\'etricos e magn\'eticos de um volume, menos a energia dissipada pelo calor, tem que ser igual \`a pot\^encia que deixa a superf\'{\i}cie fechada que limita este volume. A express\~ao \'e (assumindo um meio linear, isotr\'opico e invariante no tempo):

\begin{eqnarray}
\oint_{A} (\vec{E} \times \vec{H}) \cdot \ud \vec{A} = 
& - & \int_{V} \vec{J} \cdot \vec{E} \ud V 
- {\partial \over \partial t} \int {1 \over 2} \epsilon E ^2 \ud V - \nonumber \\
& - & {\partial \over \partial t} \int {1 \over 2} \mu H ^2 \ud V.
\end{eqnarray}
Trata-se portanto de uma express\~ao para a lei de conserva\c c\~ao da energia em eletromagnetismo, e pode ser obtida a partir das Eqs. de Maxwell. O {\sl vetor de Poynting} instant\^aneo \'e dado por:
\begin{equation}
\vec{P} \equiv \vec{E} \times \vec{H}.
\end{equation}
Representa a densidade e a dire\c c\~ao do fluxo de pot\^encia e tem unidades de [Watts][m]$^{-2}$.

\clearpage

\section{Ap\^endice}

\subsection{Sum\'ario das Leis de Maxwell}

\begin{table}[h]
\caption{\label{tab-sum-int} Leis integrais no v\'acuo}
\begin{tabular}{|c|c|} \hline
 & \\ 
Gauss            & $\oint_{A} \vec{D}^* \cdot \ud \vec{A} = \int_{V} \rho \ud {\mathrm V}$   \\ 
Amp\`ere         & $\oint_{C}\vec{H}\cdot \ud \vec{l} = \int_{A} \vec{J} \cdot \ud \vec{A}
+ {\ud \over \ud t} \int_{A} \vec{D}^* \cdot \ud \vec{A}$ \\
Faraday          & $\oint_{C}\vec{E}\cdot \ud \vec{l} = - {\ud \over \ud t}
\int_{A} \vec{B}^* \cdot \ud \vec{A}$ \\
Gauss Campo Mag. & $\oint_{A} \vec{B}^* \cdot \ud \vec{A} = 0$   \\
Cons. carga      & $\oint_{A} \vec{J} \cdot \ud \vec{A} = - {\ud \over \ud t} \int_{V} \rho \ud 
{\mathrm V}$ \\
 & \\ \hline
\end{tabular}
\end{table}

\begin{table}[h]
\caption{\label{tab-sum-dif} Leis diferenciais no v\'acuo}
\begin{tabular}{|c|c|} \hline
 & \\
Gauss            &  $\nabla \cdot \vec{D}^* = \rho$  \\ 
Amp\`ere         &  $\nabla \times \vec{H} = \vec{J} + {\partial \vec{D}^* \over \partial t}$ \\
Faraday          &  $\nabla \times \vec{E} = - {\partial \vec{B}^* \over \partial t}$ \\
Gauss Campo Mag. & $\nabla \cdot \vec{B}^* = 0$  \\
Cons. carga      & $\nabla \cdot \vec{J} = - {\partial \rho \over \partial t}$ \\
 & \\ \hline
\end{tabular}
\end{table}

Ver as defini\c c\~oes de $\vec{B}^*$ e $\vec{D}^*$ nas Eqs. \ref{B*}, \ref{D*}.

Em meios materiais, basta fazer as substitui\c c\~oes abaixo usando as Eqs. \ref{D}, \ref{B}:

\begin{eqnarray}
\vec{D}^* & \rightarrow & \vec{D} \nonumber \\
\vec{B}^* & \rightarrow & \vec{B} \nonumber \\
\rho & \rightarrow & \rho_{livre}
\end{eqnarray}

As substitui\c c\~oes correspondentes para o caso de meios lineares e isotr\'opicos devem ser feitas usando as Eqs. \ref{Dlin} e \ref{Blin}.

\vspace{2.5cm}

\subsection{Algumas f\'ormulas \'uteis}
\begin{equation}
e^{i\theta}= \cos \theta + i \sin \theta. \label{euler}
\end{equation}
\begin{equation}
\nabla \cdot (\nabla \times \vec{F}) = 0. \label{form1}
\end{equation}
\begin{equation}
\nabla \times (\nabla \Phi) = 0. \label{form2}
\end{equation}
\begin{equation}
\nabla \times (\nabla \times \vec{F}) = \nabla (\nabla \cdot \vec{F}) - \nabla^2\vec{F}, \label{form3}
\end{equation}
onde $\nabla^2 \equiv \nabla \cdot \nabla$.
\begin{equation}
\nabla \cdot \vec{r} = 3.
\end{equation}
\begin{equation}
\nabla \times \vec{r} = 0.
\end{equation}
\begin{equation}
\nabla \left ( {1 \over r} \right ) = -{1 \over r^2} \hat{r}.
\end{equation}
\begin{equation}
\nabla \cdot \left ( {1 \over r^2}  \hat{r}  \right ) = 4 \pi \delta^3(\vec{r}).
\end{equation}
\begin{equation}
\nabla^2  \left ( {1 \over r} \right ) = - 4 \pi \delta^3(\vec{r}). \label{1r}
\end{equation}

{\noindent \rule{57ex}{0.2ex}\\}
{\bf Exerc\'{\i}cio 9:} Demonstre a Eq. \ref{1r} atrav\'es da Eq. de Poisson [Eq. \ref{poisson}] para uma carga pontual.

{\sl Solu\c c\~ao:} O potencial el\'etrico de uma carga pontual \'e (mostre isso, usando as Eq. \ref{Epontual} e \ref{Egrad}):
\begin{equation}
\Phi = {q\over 4 \pi \epsilon_0 r}.
\end{equation}
A densidade de carga de uma carga pontual \'e dada por: $\rho=q\delta^3(\vec{r})$, o que nos d\'a:
\begin{equation}
\nabla^2 \Phi = \nabla^2 \left ( {q\over 4 \pi \epsilon_0 r} \right ) = - {\rho \over \epsilon_0} = 
- {q\delta^3(\vec{r}) \over \epsilon_0}.
\end{equation}

{\noindent \rule{57ex}{0.2ex}\\}


\end{document}